\documentclass[aip,sd,amsmath,amssymb,reprint]{revtex4-1}

\usepackage{graphicx}
\usepackage{dcolumn}
\usepackage{bm}
\usepackage{multirow}

\newcommand{\hyb}{$\text{sc-PBE0}\alpha_{\epsilon_{\infty}}$}
\newcommand{\al}{$\text{[AlO}_4\text{]}^0$}
\newcommand{\all}{$\text{[AlO}_4\text{]}^{-1}$}

\begin{document}

\title{Hole localization in Al-doped quartz SiO$_2$ within \emph{ab initio} hybrid-functional DFT}

\author{Matteo Gerosa}
\affiliation{Department of Energy, Politecnico di Milano, via Ponzio 34/3, 20133 Milano, Italy}
\author{Cristiana Di Valentin}
\affiliation{Dipartimento di Scienza dei Materiali, Universit\`a di Milano-Bicocca, via R. Cozzi 55, 20125 Milan, Italy}
\author{Carlo Enrico Bottani}
\email[Corresponding author: ]{carlo.bottani@polimi.it}
\affiliation{Department of Energy, Politecnico di Milano, via Ponzio 34/3, 20133 Milano, Italy}
\affiliation{Center for Nano Science and Technology @Polimi, Istituto Italiano di Tecnologia, via Pascoli 70/3, 20133 Milano, Italy}
\author{Giovanni Onida}
\affiliation{Dipartimento di Fisica, Universit\`a degli Studi di Milano, Via Celoria 16, Milano, Italy}
\affiliation{European Theoretical Spectroscopy Facility (ETSF)}
\author{Gianfranco Pacchioni}
\affiliation{Dipartimento di Scienza dei Materiali, Universit\`a di Milano-Bicocca, via R. Cozzi 55, 20125 Milan, Italy}

\date{\today}

\begin{abstract}

We investigate the long-standing problem of the hole localization at the Al impurity in quartz SiO$_2$, using a relatively recent DFT hybrid-functional method in which the exchange fraction is obtained \emph{ab initio}, based on an analogy with the static many-body COHSEX approximation to the electron self-energy. As the amount of the admixed exact exchange in hybrid functionals has been shown to be determinant for properly capturing the hole localization, this problem constitutes a prototypical benchmark for the accuracy of the method, allowing one to assess to what extent self-interaction effects are avoided. We obtain good results in terms of description of the charge localization and structural distortion around the Al center, improving with respect to the more popular B3LYP hybrid-functional approach. We also discuss the accuracy of computed hyperfine parameters, by comparison with previous calculations based on other self-interaction-free methods, as well as experimental values. We discuss and rationalize the limitations of our approach in computing defect-related excitation energies in low-dielectric-constant insulators.

\end{abstract}

\maketitle

\section{Introduction}

The Al impurity is one of the most commonly observed defects in irradiated quartz SiO$_2$, by which a tetravalent Si cation is replaced with a trivalent Al atom. The Al/Si substitution results in an unpaired electron (or, in an equivalent description, a hole) which, based on early experimental observations, has been identified to be trapped in a nonbonding $2p$ orbital of an O atom surrounding the substitutional Al.\cite{griffiths1954,obrien1955} The corresponding neutral Al defect center (hereafter also denoted \al) is magnetically active, and has been the subject of extensive characterization by electron paramagnetic resonance (EPR),\cite{nuttal1980,nuttal1981,schnadt1971} also in combination with absorption spectroscopy.\cite{nassau1975,meyer1984,schirmer1976}

From the theory side, it has been recognized that the reproduction of the experimentally observed features of the \al center is subordinate to a correct description of the hole localization properties. Since early density-functional theory (DFT) calculations using (semi)local density functionals gave a wrong picture, predicting the hole to be delocalized over all the four Al nearest-neighbor O atoms,\cite{continenza1996,magagnini2000,laegsgaard2000} this problem has been identified as a challenging testing ground for novel density-functional methods.\cite{laegsgaard2001}

The wrong description of (semi)local DFT functionals has been ascribed to incomplete cancellation of the self-interaction (SI) brought in by the Hartree term,\cite{laegsgaard2000,pacchioni2000} since calculations using exactly SI-free Hamiltonians, such as unrestricted Hartree-Fock (UHF) and self-interaction-corrected (SIC) DFT, yield the correct hole localization. Pacchioni~\emph{et al.} also performed unrestricted second-order M\o{}ller-Plesset perturbation theory (UMP2) calculations,\cite{pacchioni2000} concluding that no appreciable role is played by correlation in the specific problem at hand.

A now routinely used approach allowing to partially amend the SI error is based on admixing a fraction of Hartree-Fock exact exchange (EXX) to semilocal exchange-correlation functionals. However, popular recipes for such hybrid functionals, prescribing $20\%$ or $25\%$ of EXX, such as in B3LYP,\cite{becke1993b,stephens1994} PBE0,\cite{perdew1996} and HSE06 (Ref.~\onlinecite{heyd2003,* heyd2006}) have proven to fail in describing charge localization in Al-doped silica.\cite{laegsgaard2001,gillen2012,pacchioni2000, solans-monfort2004,zyubin2003} Thus, it has been argued that a large amount of EXX would be necessary to obtain agreement with experiments,\cite{laegsgaard2001,solans-monfort2004} much like a large enough Hubbard $U$ parameter is needed within DFT+$U$ to solve the same problem.\cite{nolan2006} For instance, To~\emph{et al.} found that a semi-empirical hybrid functional including $42\%$ of EXX (called BB1K functional) yields the correct picture.\cite{to2005}

In view of the preceding work, it may be concluded that some amount of empiricism is required to tackle the Al-impurity problem within DFT, casting doubts on its actual predictive power. Recently, a rationale for the value that the EXX fraction takes in extended systems has been put forward.\cite{alkauskas2011,marques2011} Starting from the many-body $GW$ approximation to the electron self-energy $\Sigma(\omega)$, one obtains, in its static limit ($\omega \to 0$), the so-called Coulomb-hole-plus-screened-exchange (COHSEX) approximation to $\Sigma$.\cite{hedin1965} By performing a spatial average of the electron gas polarization function, one is finally left with a simple analytical expression for the EXX fraction, which is now expressed in terms of the macroscopic electronic dielectric constant of the material. Thus, the EXX fraction can be obtained \emph{ab initio} for a given material once its dielectric constant is computed within DFT; a self-consistent approach to the definition of such \emph{self-consistent dielectric-dependent hybrid functional} has been recently proposed and tested for various properties of oxide semiconductors and insulators.\cite{skone2014,gerosa2015} Notice that COHSEX treats the exchange term exactly, and hence is SI free; it thus constitutes the ideal candidate for modeling systems whose ground state is dominated by the classical Hartree and electron exchange Fock interactions.

\section{Computational Details}

All-electron DFT calculations were performed within the linear combination of atomic orbitals approach as implemented in the \textsc{crystal09} code.\cite{crystal09-manual,dovesi2005} The following all-electron basis sets were employed: 66-21G$^{*}$ for Si (Ref.~\onlinecite{nada1990}), 8-411($d$1) for O (Ref.~\onlinecite{ruiz2003}), 85-11G$^{*}$ for Al (Ref.~\onlinecite{catti1994}). A full-range hybrid functional\cite{becke1993} was adopted for the treatment of exchange and correlation; the fraction of admixed EXX ($\alpha$) was evaluated based on the above-mentioned relationship with the macroscopic electronic dielectric constant ($\epsilon_{\infty}$), $\alpha \approx 1/\epsilon_{\infty} $,\cite{alkauskas2011,marques2011} and obtained self-consistently for pristine quartz SiO$_2$, following the procedure illustrated in Ref.~\onlinecite{gerosa2015}. The resulting functional is referred to as ``\hyb'' in the following, as it is \emph{de facto} a self-consistent (sc) re-parametrization of the PBE0 hybrid functional.\cite{perdew1996}
The dielectric constant $\epsilon_{\infty}$ was computed within the coupled-perturbed Kohn-Sham method implemented in the \textsc{crystal09} code.\cite{ferrero2008a,ferrero2008b}

The Al center was modeled in an embedding $2\times2\times2$ quartz SiO$_2$ supercell (72 atoms) with the atomic positions and lattice parameters fully relaxed for the bulk cell using the \hyb functional. For the defective supercell, further optimization of the atomic positions was carried out at fixed lattice parameters. The standard thresholds defined in \textsc{crystal09} were adopted in all geometry optimizations.\cite{Note1} The Brillouin zone was sampled by using 8 $k$-points in the irreducible wedge.\cite{monkhorst1976}

The charge-transition levels were computed according to the formalism illustrated in Refs.~\onlinecite{vandewalle2004,lany2008}. In particular, total-energy differences relative to the defect charge state variation were computed using defect Kohn-Sham (KS) eigenvalues, following the approach proposed in Ref.~\onlinecite{gallino2010}. The $1s$ KS eigenvalue of Si was taken as reference for aligning band structures in defect and bulk calculations. The spurious electrostatic interaction between image charged defects was accounted for by correcting the KS eigenvalues according to the procedure illustrated by Chen and Pasquarello,\cite{chen2013} and based on the Makov-Payne correction scheme.\cite{leslie1985,makov1995}

\section{Results and discussion}

Within the \hyb approach, the dielectric constant of quartz SiO$_2$ turns out to be $\epsilon_{\infty} = 2.15$ (the experimental value is 2.38, see Ref.~\onlinecite{chang2000} and references therein), which corresponds to an exchange fraction $\alpha = 46.5\%$. The method predicts a somewhat overestimated band gap of 11.6~eV, whereas various experiments measured it in the quite broad range of $8-10$~eV.\cite{weinberg1979} The failure of the \hyb functional in computing reliable band gaps for some insulators with very low dielectric constants (and correspondingly large band gaps) was already reported in Refs.~\onlinecite{skone2014,gerosa2015}, being particularly serious when the geometry is re-optimized at each self-consistency step (see for instance the case of MgO, which has a dielectric constant of $\sim 3$).\cite{gerosa2015} 

However, here we are mainly concerned with the description of the hole localization at the \al center, which is a ground state property; the related spectroscopic features will be deferred to a separate discussion in Section~\ref{subs:optical_properties}. 

\subsection{Structural deformation and hole localization}

\begin{figure}[t]
 \includegraphics{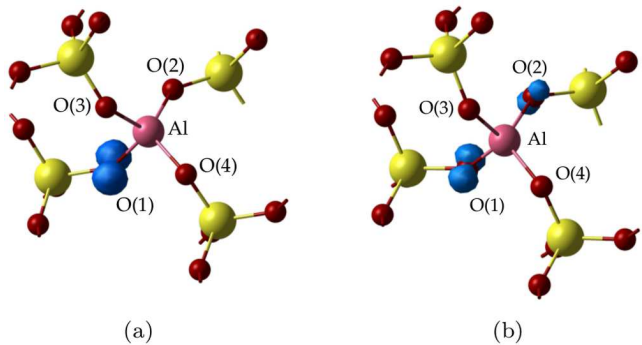}
 \caption{\label{fig1} Ball-and-stick representation (Si, O and Al atoms are shown as yellow, red, and pink spheres, respectively) of the local atomic structure around the Al impurity, as obtained from geometry optimization at the (a) \hyb, and (b) B3LYP level of theory. Isosurface of the spin density associated with the unpaired electron introduced by the \al center is shown. For bond distances see Table~\ref{tab1}.}
\end{figure}

Figure~\ref{fig1}(a) shows the local atomic structure of SiO$_2$ around the Al impurity, as found by minimizing the total electronic energy using the \hyb functional and allowing for symmetry-breaking atomic relaxations, which amounts at independently optimizing the positions of each atom of the (O$_1$, O$_2$) and (O$_3$, O$_4$) oxygen pairs (where the two O within each pair are equivalent to each other in the bulk SiO$_2$ structure). The emerging picture agrees very well with that obtained within other rigorously SI-free approaches, such as UHF\cite{pacchioni2000,laegsgaard2001} and SIC-DFT.\cite{davezac2005} In particular, the hole introduced by the substitutional Al atom is trapped at the O(1) atom, its wavefunction exhibiting purely $2p$ character, with the corresponding orbital lying almost perpendicularly to the Al-O-Si plane. As a consequence of the charge localization, the local atomic structure distorts considerably: the O(1) atom moves away from the Al atom, resulting in an average equilibrium Al-O distance $13\%$ larger compared to the other Al-O distances (see also Table~\ref{tab1}).

The results of our calculations are in agreement with the experimental evidence as obtained from EPR investigations:\cite{nuttal1981} (i) the hole localizes in the nonbonding $2p$ orbital of the O atom corresponding to the longer Si-O-type bond in pure SiO$_2$; (ii) the $2p$ orbital is perpendicular to the Al-O-Si plane; (iii) the localization causes the hole-bearing oxygen to move away $12\%$ farther from the Al center with respect to the other O atoms.

For comparison, we also tested the performance of the Becke's three-parameter hybrid functional (B3LYP),\cite{becke1993b,stephens1994} incorporating $20\%$ of EXX. From Table~\ref{tab1} it is inferred that, while some asymmetry is still present in the resulting optimized structure, the elongation of the Al-O(1) distance is now at most $6\%$ larger than the other Al-O distances. The Mulliken population analysis presented in Table~\ref{tab2} shows that in the B3LYP ground state the unpaired electron charge density is distributed over the (O(1), O(2)) pair, with a substantial contribution also from a $2p$ orbital of the O(2) atom [see also Figure~\ref{fig1}(b)], at variance with the experimental evidence. 
The final picture is not dissimilar if the optimization is started from the distorted structure of the \al center optimized at the \hyb level:\cite{Note2} the Al-O(1) elongation now amounts at about $7\%$, and the unpaired electron is delocalized over the (O(1), O(2)) pair with roughly the same proportions as reported in Table~\ref{tab2}. 
In some of the previous B3LYP studies of Al-doped SiO$_2$ the inequivalence of the O(1) and O(2) sites is not even qualitatively captured (same Al-O distances), and the hole is found evenly delocalized over two, or even over all the four O atoms nearest to the Al impurity.\cite{pacchioni2000, laegsgaard2001}

\begin{table}[t]
\caption{\label{tab1} Nearest-neighbor Si-O and Al-O distances ($\text{\AA}$) for pure bulk and Al-doped SiO$_2$ (\al), respectively. The O atoms are labeled according to Figure~\ref{fig1}.}
\begin{ruledtabular}
\begin{tabular}{lcccc}
 & Pure SiO$_2$ & \multicolumn{3}{c}{\al} \\
 \cline{3-5}
Functional & \hyb & \hyb 
& B3LYP\footnote{Starting geometry for optimization: pure SiO$_2$ structure.} & B3LYP\footnote{Starting geometry for optimization: \hyb-optimized \al structure.} \\
\hline
O(1) & 1.620 & 1.910 & 1.809 & 1.826 \\
O(2) & 1.620 & 1.699 & 1.758 & 1.749 \\
O(3) & 1.616 & 1.687 & 1.705 & 1.705 \\
O(4) & 1.616 & 1.691 & 1.700 & 1.700 \\
\end{tabular}
\end{ruledtabular}
\end{table}

\begin{table}[t]
\caption{\label{tab2} Spin population of the O atoms belonging to the \al center, and EPR hyperfine parameters of the hole-bearing $^{17}$O and of the $^{27}$Al. The optimized structures are obtained starting from ideal bulk SiO$_2$.}
\begin{ruledtabular}
\begin{tabular}{lcccc}
Functional & \hyb & B3LYP & B3LYP & Expt. \\
Geometry & \hyb & B3LYP & \hyb & (Refs.~\onlinecite{nuttal1981,nuttal1980}) \\
\hline
\multicolumn{5}{c}{Spin population} \\
O(1) & 0.95 & 0.58 & 0.81 \\ 
O(2) & 0.02 & 0.28 & 0.07 \\ 
O(3) & $<0.01$ & 0.01 & 0.03 \\
O(4) & $<0.01$ & 0.04 & $<0.01$ \\
\\
\multicolumn{5}{c}{$^{17}$O(1) hyperfine matrix (G)\footnote{The principal values of the anisotropic hyperfine matrix are listed so that $B_1 < B_2 < B_3$.}} \\
$A_{\text{iso}}$ & -42.6 & -26.1 & -34.7 & -26.0 \\
$B_1$            & -94.3 & -61.7 & -83.7 & -85.0 \\
$B_2$            &  47.1 &  30.7 &  41.8 &  41.2 \\
$B_3$            &  47.2 &  31.0 &  41.9 &  43.8 \\
\\
\multicolumn{5}{c}{$^{27}$Al hyperfine matrix (G)$^{\text{a}}$} \\
$A_{\text{iso}}$ & -5.0 & -8.1 & -5.4 & -5.8 \\
$B_1$            & -0.4 & -0.2 & -0.4 & -0.4 \\
$B_2$            & -0.4 & -0.1 & -0.3 & -0.3 \\
$B_3$            &  0.8 &  0.3 &  0.6 &  0.7 \\
\end{tabular}
\end{ruledtabular}
\end{table}

\subsection{EPR parameters}

In order to further confirm the better performance of the \hyb approach with respect to B3LYP, we computed hyperfine parameters relative to the hole-bearing O(1) atom. The hyperfine coupling matrix, describing the magnetic interaction of the spin of the unpaired electron with the spin of the neighboring nuclei ($^{17}$O and $^{27}$Al), is conveniently divided into an isotropic (spherically symmetric) and an anisotropic (dipolar) part. The isotropic part (denoted $A_{\text{iso}}$) is proportional to the electron spin density at the nucleus, and, as such, the dominant contribution to it is caused by spin-polarization of the $s$ electrons. Instead, the anisotropic part is related to the spin population of orbitals with higher angular momentum components; this contribution is commonly expressed in terms of a matrix with principal values $B_1$, $B_2$ and $B_3$ reported in Table~\ref{tab2}. Since the hole wavefunction has purely $2p$ character, the dipolar part gives direct access to the corresponding spin distribution. Instead, the isotropic contribution is notably harder to be reproduced, being extremely sensitive to the details of the calculation in general, and to the choice of the basis set in particular.\cite{barone1996} We nevertheless report on it as well for the sake of completeness.

\begin{table*}[tb]
\caption{\label{tab3} Vertical excitation energies (in eV) associated to the \al center computed with the \hyb and B3LYP functionals at different optimized geometries. Comparison with theoretical results from the literature obtained within time-dependent DFT (TDDFT) using the B3LYP and BB1K exchange-correlation approximations, as well as within the outer-valence Green's function (OVGF) approach. Experimental position of the main absorption peak is also reported.}
\begin{ruledtabular}
\begin{tabular}{lcccccc}
\multicolumn{3}{c}{This work} \\
\cline{1-3}
Functional & \hyb & B3LYP & TDDFT-BB1K & TDDFT-B3LYP & OVGF & Expt. \\
Geometry & \hyb & \hyb & (Ref.~\onlinecite{to2005}) & (Ref.~\onlinecite{zyubin2003}) & (Ref.~\onlinecite{zyubin2003}) & (Ref.~\onlinecite{nassau1975}, Ref.\onlinecite{meyer1984}) \\
\hline
& 4.73 & 2.91 & 3.03 & 1.72 & 2.74 & 2.9, 2.85 \\
\end{tabular}
\end{ruledtabular}
\end{table*}

In Table~\ref{tab2} the hyperfine parameters are reported for the $^{17}$O(1) and $^{27}$Al nuclei. Concerning the anisotropic parameters of $^{17}$O, our \hyb calculations nicely capture the experimentally observed strong anisotropy along the three axes, and numerical values are in good agreement with both experiments and the results of previous investigations based on SI-free approaches (UHF, UMP2, SIC-DFT).\cite{pacchioni2000,davezac2005,laegsgaard2001} In contrast, B3LYP yields a quantitatively wrong picture, the computed parameters being substantially smaller in absolute value than experimental ones. The situation is quite the opposite for the isotropic part, for which \hyb overestimates the absolute value of $A_{\text{iso}}$, while B3LYP yields it exceptionally close to experiment. We argue that the latter result is fortuitous, in the sense that it is not concomitant with a correspondingly more accurate description of the hole localization. Firstly, as already discussed, at the B3LYP level the O(2) atom carries a substantial part of the hole-related spin density (see Table~\ref{tab2}), in disagreement with experiment, and accordingly the $^{17}$O(2) EPR parameters are of the same order of magnitude as for $^{17}$O(1);\cite{Note3} instead, when the hole localization is correctly captured, such as at the \hyb level, the former are at least an order of magnitude smaller than the latter (see also Ref.~\onlinecite{to2005}). Secondly, the improvement of the isotropic part does not come along with a similar improvement of the anisotropic one, which indeed is related to the proper description of the $2p$ hole wavefunction.

The above conclusion is also supported by the computed superhyperfine matrix of $^{27}$Al: the \hyb functional gives results in quantitative agreement with both experiment and UHF calculations,\cite{pacchioni2000} whereas this is not true for B3LYP. Notice that the superhyperfine interaction with a dopant element like Al that introduces a hole in the structure is usually the only accessible information. In fact, in order to measure the O hyperfine constants, $^{17}$O enriched samples have to be prepared with complex and costly procedures.\cite{gionco2015}

\subsection{Optical properties of the Al impurity}
\label{subs:optical_properties}

The Al impurity in quartz silica has been observed to act as a color center, endowing it with the typical smoky coloration. However, considerable controversy arose as to which absorption feature had to be correlated with such observation.\cite{schirmer1976,nassau1975} It was finally concluded that an absorption peak at about 2.9~eV is to be associated with the presence of Al centers and, thus, with the smoky coloring.\cite{meyer1984} 

The observed optical transition should be related to excitation of the hole trapped at the neutral \al center ($q=0$) into the VB, leading to a negatively charged defect ($q'=-1$, \all center); accordingly, from the theory side, the optical transition level $(0/-1)$ is the relevant quantity to be compared with experiments. The \hyb predicts the computed level to be $\sim4.7$~eV above the valence band (VB) maximum, nearly 2~eV higher in the band gap than in experiments. We attribute this disagreement to the already mentioned overestimation of the electronic band gap of bulk quartz provided by \hyb. Consequently, the Al-related defect level is wrongly positioned with respect to the VB ($\sim 5$~eV above its edge) and this eventually gives rise to the observed overestimation of the optical transition energy. 

As a partial workaround, we computed the electronic structure using the B3LYP functional, which yields a band gap of 8.6~eV for bulk quartz, falling in the range of the experimental values; the \al geometry obtained within \hyb was instead retained. The spin population and EPR parameters computed following this approach are reported in Table~\ref{tab2}. The hole is again localized on the O(1) atom, although with a Mulliken density lower than that obtained by performing calculations fully within the \hyb scheme. Surprisingly, the computed hyperfine parameters are in even better quantitative agreement with experiments, as the lower spin density counterbalances the overestimation yielded by the \hyb; a similar trend was noticed in the previous hybrid-functional investigation of To~\emph{et al.}\cite{to2005} The charge density distribution analysis for the negatively-charged center led us to conclude that the same qualitative picture is obtained at the \hyb and B3LYP levels, provided that the geometry is kept fixed to the one optimized within \hyb: the one-particle state corresponding to vertical excitation of the hole to the VB is still contributed by the $2p$ orbitals of the O(1) atom. The computed optical level for such transition is positioned at $\sim2.9$~eV above the top of the VB, in excellent agreement with experiment. For comparison, we report in Table~\ref{tab3} representative results from previous theoretical studies for the computed vertical transition energy correlating with the experimentally found absorption band with a maximum at $\sim2.9$~eV.

\section{Conclusions}

We have re-investigated the long-standing problem of hole localization in Al-doped quartz SiO$_2$ using a recently proposed hybrid-functional method in which the exchange fraction is consistently determined based on the analogy with the many-body COHSEX approximation to the electron self-energy, without empirical prescriptions or fitting to the experimental data. The COHSEX scheme is rigorously SI-free, and thus constitutes the ideal starting point for studying systems in which incomplete cancellation of SI leads to a qualitatively wrong ground state. The neutral Al impurity in quartz silica is just a paradigmatic case: (semi)local or standard hybrid DFT functionals fail in capturing the experimentally evidenced hole localization at one of the Al-coordinated O atoms.\cite{laegsgaard2001,pacchioni2000} In particular, the failure of popular hybrid functionals, such as B3LYP, has been attributed to the insufficient amount of EXX admixed.

The \hyb approach allows one to evaluate the exchange fraction from first-principles, based on a simple relationship with the macroscopic dielectric constant of the material. In the case of quartz silica, we obtained it to be $\sim 46\%$. The resulting hybrid functional correctly reproduces the hole localization at a single O atom surrounding the Al impurity, also giving an accurate description of the structural distortion around it, and allowing to compute EPR parameters in agreement with previous SI-free calculations\cite{laegsgaard2001,pacchioni2000,to2005} and experiments.\cite{nuttal1981} However, the defect-related optical spectroscopic features are not well-reproduced by the \hyb method. We attribute this failure to the observed overestimation of the bulk quartz silica band gap. Using B3LYP on top of the \hyb optimized geometry yields a band structure in better agreement with experiment and corroborates this hypothesis.

In conclusion, we have shown that the hybrid-functional approach tested in this work is capable of correcting most of the SI error inherent to (semi)local, as well as more popular hybrid, DFT functionals. This feature is crucial for adequately describing the ground state of defective oxide materials. As far as defect-related excitation energies are concerned, their determination is subordinate to an accurate calculation of the electronic structure of the bulk material. This is not always the case for low-dielectric-constant insulators in general, and for quartz SiO$_2$ in particular. However, based on our experience, the \hyb method is able to reproduce the whole experimental scenario when point defects in moderate gap metal-oxide semiconductors (dielectric constants $\sim 4-6$) are addressed.

\begin{acknowledgments}

This work has been supported by the Italian MIUR through the FIRB Project RBAP115AYN ``Oxides at the nanoscale: multifunctionality and applications''.
The support of the COST Action CM1104 “Reducible oxide chemistry, structure and functions” is also gratefully acknowledged. G.O. acknowledges
the ETSF-Italy\cite{etsf} for computational support.

\end{acknowledgments}

\end{document}